# Micromechanics and theory of point defects in anisotropic elasticity: Eshelby factor meets Eshelby tensor

Markus Lazar*

*Department of Physics*
*Darmstadt University of Technology*
*Hochschulstr. 6, D-64289 Darmstadt, Germany*

**Abstract** The interaction of anisotropic point defects in anisotropic media is studied in the framework of anisotropic elasticity with eigendistortion. For this purpose key-equations and their solutions for anisotropic point defects in an anisotropic medium based on the anisotropic Green tensor are derived. The material force, interaction energy and torque between two point defects as well as between a point defect and a dislocation loop are given. We discuss so-called contact terms and point out similarities between elastic, electric, and magnetic dipoles. The plastic, the elastic and the total volume changes caused by an anisotropic point defect in an anisotropic material and the related Eshelby factor are determined. Thereby, the Eshelby factor is given in terms of the Eshelby tensor.

## 1. Introduction

Point defects are important defects in crystals [Leibfried and Breuer, 1978; Balluffi and Granato, 1979; Teodosiu, 1982; Balluffi, 2012]. Point defects in crystals can exist in different configurations such as vacancies, interstitials, and substitutional atoms. The fields of point defects play an important role in determining the physical properties of solids. They cause volume change and interact with dislocations if dislocations climb. Dislocations only climb if point defects can migrate to or from them. Especially, point defects play a major role in many physical problems such as X-ray scattering, internal friction phenomena, aggregation of defects, dislocation locking and various diffusion processes [Nabarro, 1967]. Thus, the elastic interaction between point defects as well as between point defects and dislocations are important in defect mechanics. Besides elastic interaction, the interaction between point defects and dislocations comes from other fields, such as chemical (Suzuki-type interaction), electrical and geometric (interaction between dislocation core and point defect) aspects (see, e.g., Teodosiu [1982]; Hull and Bacon [2011]). An important property of any point defect is its symmetry, which can be either the same as the symmetry of the host crystal or lower (see, e.g., Balluffi [2012]; Teodosiu [1982]). Micromechanics of defects based on Green tensors is an active and important research field of the so-called eigenstrain theory [Mura, 1987; Buryachenko, 2007; Li and Wang, 2008]. In such a framework, point defects can be modeled as Dirac $\delta$-type singularities in the eigendistortion [deWit, 1973b]. Point defects cause self-stresses in crystals and can be extrinsic and intrinsic defects [Kröner, 1960]. Nowadays, the fields of point defects are important for computer simulations of defect mechanics, especially for a realistic modeling of dislocation climb in dislocation dynamics.

* *E-mail address:* lazar@fkp.tu-darmstadt.de (M. Lazar).

The paper is organized into five sections. In Sec. 2, anisotropic elasticity with eigendistortion, the elastic Green tensor, the first and second derivatives of the Green tensor, related kernel functions, and key-equations for arbitrary eigendistortion are given. In Sec. 3, anisotropic point defects in an anisotropic medium are studied. The corresponding solutions for the displacement vector, elastic distortion tensor, total distortion tensor and stress tensor are derived. In addition, the material force, interaction energy and elastic torque between two point defects as well as between a point defect and a dislocation loop are given. The explicite form of these expressions in terms of the gradients of the Green tensor and contact terms is a new contribution to the theory of point defects and is given in this work for the first time. In Sec. 4, the Eshelby factor is given in terms of the Eshelby tensor for anisotropic materials. In Sec. 5, conclusions are presented.

## 2. Eigendistortion in anisotropic elasticity

In this Section, we derive key-equations for arbitrary eigendistortion in an anisotropic material using the eigenstrain theory. We are using the Green tensor method since it provides a convenient framework to calculate the fields of point defects in anisotropic media.

### 2.1. *Basics*

In the theory of anisotropic elasticity, the static equilibrium condition for self-stresses (no body forces) reads

$$\sigma_{ij,j} = 0\,, \tag{1}$$

where $\boldsymbol{\sigma}$ is the symmetric stress tensor. We use the indicial comma notation to indicate spatial differentiation: $\partial_j$ is indicated by the subscript notation ",$j$". The stress tensor is related to the elastic distortion tensor $\boldsymbol{\beta}$ by Hooke's law

$$\sigma_{ij} = C_{ijkl}\beta_{kl}\,, \tag{2}$$

where $C_{ijkl}$ is the fourth rank tensor of elastic constants. The tensor $C_{ijkl}$ possesses the so-called minor symmetries

$$C_{ijkl} = C_{jikl} = C_{ijlk} \tag{3}$$

and the so-called major symmetry

$$C_{ijkl} = C_{klij} \,. \tag{4}$$

Therefore, the tensor $C_{ijkl}$ has 21 independent components for arbitrary elastic anisotropy.

The gradient of the displacement vector $\boldsymbol{u}$, which is the total distortion tensor $\boldsymbol{\beta}^{\mathrm{T}}$, can be decomposed into an elastic distortion tensor $\boldsymbol{\beta}$ and a plastic (or anelastic) distortion tensor $\boldsymbol{\beta}^{\mathrm{P}}$

$$\beta_{ij}^{\mathrm{T}} := u_{i,j} = \beta_{ij} + \beta_{ij}^{\mathrm{P}} \,. \tag{5}$$

In micromechanics, the plastic distortion tensor plays the physical role of an eigendistortion tensor being a stress-free distortion. For point defects, the eigendistortion tensor is often called quasi-plastic distortion tensor [Kröner, 1960; deWit, 1981]. The quasi-plastic distortion may serve as a defect density in its own right. If the eigendistortion tensor or quasi-plastic distortion tensor is non-constant (or non-uniform), the corresponding dislocation density tensor, which may be called the quasi-dislocation density tensor, is defined by [Kröner, 1958, 1960; deWit, 1981]

$$\alpha_{ij} = -\epsilon_{jkl}\beta_{il,k}^{\mathrm{P}} \,, \tag{6}$$

where $\epsilon_{jkl}$ denotes the Levi-Civita tensor. As mentioned by Kröner [1956, 1958, 1981] and deWit [1973b] the plastic distortion $\boldsymbol{\beta}^{\mathrm{P}}$ of a point defect can be regarded as equivalent to an "infinitesimal dislocation loop density" being a fictitious dislocation distribution. The dislocation density tensor (6) satisfies the Bianchi identity

$$\alpha_{ij,j} = 0 \,. \tag{7}$$

## 2.2. *Key-equations for eigendistortions*

Substituting Eqs. (2) and (5) into Eq. (1), we obtain an inhomogeneous Navier equation for the displacement vector

$$C_{ijkl}u_{k,lj} = C_{ijkl}\beta_{kl,j}^{\mathrm{P}} \,, \tag{8}$$

where the inhomogeneous part (or source part) on the right hand site is given by the gradient of the eigendistortion or quasi-plastic distortion tensor $\boldsymbol{\beta}^{\mathrm{P}}$. The Green tensor $G_{km}$ of the Navier equation is defined by

$$C_{ijkl}G_{km,lj}(\boldsymbol{R}) + \delta_{im}\delta(\boldsymbol{R}) = 0 \,. \tag{9}$$

Here, $\delta_{im}$ denotes the Kronecker delta tensor and $\delta(\boldsymbol{R})$ is the three-dimensional Dirac delta function. If we use the Green tensor, then the solution of Eq. (8) is given by the convolution of the (negative) Green tensor with the right hand side of Eq. (8)

$$u_i = -C_{jkln}G_{ij} * \beta_{ln,k}^{\mathrm{P}} = -C_{jkln}G_{ij,k} * \beta_{ln}^{\mathrm{P}} \,, \tag{10}$$

where $*$ denotes the spatial convolution. Eq. (10) is the generalized Volterra formula for an arbitrary eigendistortion.

The gradient of Eq. (10) delivers the total distortion

$$\beta_{im}^{\mathrm{T}} = u_{i,m} = -C_{jkln} G_{ij,km} * \beta_{ln}^{\mathrm{P}} \,. \tag{11}$$

Using Eq. (11), the elastic distortion tensor is obtained from Eq. (5) as

$$\beta_{im} = -C_{jkln} G_{ij,km} * \beta_{ln}^{\mathrm{P}} - \beta_{im}^{\mathrm{P}} \,. \tag{12}$$

If we define the kernel $R_{imln}$ [Simmons and Bullough, 1970; Lazar, 2016]

$$R_{imln}(\boldsymbol{R}) = C_{jkln} G_{ij,km}(\boldsymbol{R}) + \delta_{il}\delta_{mn}\delta(\boldsymbol{R}) \,, \tag{13}$$

then Eq. (12) reduces to

$$\beta_{im} = -R_{imln} * \beta_{ln}^{\mathrm{P}} \,. \tag{14}$$

Using Eq. (2), the stress tensor reads

$$\sigma_{pq} = -C_{pqim}\big(C_{jkln} G_{ij,km} * \beta_{ln}^{\mathrm{P}} + \beta_{im}^{\mathrm{P}}\big) \,, \tag{15}$$

which can be rewritten as

$$\sigma_{pq} = -S_{pqln} * \beta_{ln}^{\mathrm{P}} \,, \tag{16}$$

where the kernel $S_{pqln}$ is defined by [Simmons and Bullough, 1970; Kunin, 1983; Lazar, 2016]

$$\begin{aligned}
S_{pqln}(\boldsymbol{R}) &= C_{pqim} R_{imln}(\boldsymbol{R}) \\
&= C_{pqim}\big[C_{jkln} G_{ij,km}(\boldsymbol{R}) + \delta_{il}\delta_{mn}\delta(\boldsymbol{R})\big] \,.
\end{aligned} \tag{17}$$

Note that Eqs. (12) and (15) give the elastic distortion and the stress, respectively, for a prescribed eigendistortion. In micromechanics, the second derivative of the Green tensor plays an important role.

### 2.3. *Green tensor, derivatives of the Green tensor and kernels for arbitrary anisotropic elasticity*

In the theory of anisotropic elasticity, the Green tensor is given by [Lifshitz and Rosenzweig, 1947; Synge, 1957; Barnett, 1972]

$$G_{ij}(\boldsymbol{R}) = \frac{1}{8\pi^2 R} \int_0^{2\pi} (nCn)_{ij}^{-1} \, \mathrm{d}\phi \,, \tag{18}$$

where $\boldsymbol{n} = \boldsymbol{\kappa}(\pi/2, \phi)$, $\boldsymbol{R} = \boldsymbol{x} - \boldsymbol{x}'$, $R = |\boldsymbol{R}| = |\boldsymbol{x} - \boldsymbol{x}'|$ and the second rank symmetric tensor, which is the Christoffel stiffness tensor,

$$(nCn)_{ij} = n_k C_{ikjl} n_l \,. \tag{19}$$

Here $\boldsymbol{\kappa} = \boldsymbol{k}/k$ with $k = |\boldsymbol{k}|$ is a unit vector in the Fourier space with $\boldsymbol{\kappa} = \boldsymbol{\kappa}(\theta, \phi)$. Eq. (18) is the famous Lifshitz-Rosenzweig-Synge-Barnett representation of the Green tensor for arbitrary anisotropy as line integral along the unit circle in the plane orthogonal to $\boldsymbol{R}$. Thus, $\boldsymbol{n}$ is perpendicular to $\boldsymbol{R}$, namely $\boldsymbol{n} \cdot \boldsymbol{R} = 0$.

The first derivative (first gradient) of the Green tensor (18) reads [Barnett, 1972; Bacon *et al.*, 1979; Lazar, 2016]

$$G_{ij,k}(\boldsymbol{R}) = -\frac{1}{8\pi^2 R^2} \int_0^{2\pi} \left[ \tau_k (nCn)_{ij}^{-1} - n_k F_{ij} \right] \mathrm{d}\phi \qquad (20)$$

with the unit vector

$$\boldsymbol{\tau} = \frac{\boldsymbol{R}}{R} \qquad (21)$$

and

$$F_{ij} = (nCn)_{ip}^{-1} \left[ (nC\tau)_{pq} + (\tau Cn)_{pq} \right] (nCn)_{qj}^{-1} . \qquad (22)$$

On the other hand, the second derivative (second gradient) of the Green tensor can be decomposed into two terms, namely [Kunin, 1983; Kröner, 1990; Lazar, 2016]

$$G_{ij,km}(\boldsymbol{R}) = -\delta(\boldsymbol{R}) E_{ijkm} + \frac{1}{R^3} H_{ijkm} . \qquad (23)$$

Sometimes the second gradient of a Green function is called modified Green function[a] (e.g., Kröner [1990]). The first term in Eq. (23) is due to the derivative in the sense of generalized functions [Gel'fand and Shilov, 1964], and the second term is due to the formal (or ordinary) derivative of $G_{ij}$. The tensors $E_{ijkm}$ and $H_{ijkm}$ read [Lazar, 2016]

$$E_{ijkm} = \frac{1}{2\pi} \int_0^{2\pi} n_m n_k (nCn)_{ij}^{-1} \mathrm{d}\phi \qquad (24)$$

and

$$H_{ijkm} = \frac{1}{8\pi^2} \int_0^{2\pi} \left[ 2\tau_m \tau_k (nCn)_{ij}^{-1} - 2(n_m \tau_k + n_k \tau_m) F_{ij} + n_m n_k A_{ij} \right] \mathrm{d}\phi \qquad (25)$$

with

$$A_{ij} = F_{ip} \left[ (nC\tau)_{pq} + (\tau Cn)_{pq} \right] (nCn)_{qj}^{-1} + (nCn)_{ip}^{-1} \left[ (nC\tau)_{pq} + (\tau Cn)_{pq} \right] F_{qj}$$
$$- 2(nCn)_{ip}^{-1} (\tau C\tau)_{pq} (nCn)_{qj}^{-1} . \qquad (26)$$

Both tensors (24) and (25) are line integrals along the unit circle in the plane orthogonal to $\boldsymbol{R}$. The tensors (24) and (25) possess the symmetries

$$E_{ijkm} = E_{jikm} = E_{ijmk} \qquad (27)$$

$$H_{ijkm} = H_{jikm} = H_{ijmk} . \qquad (28)$$

---

[a]In the sense of generalized functions, Eq. (23) is a generalization of the so-called Frahm formula [Frahm, 1983; Kanwal, 2004]

$$\partial_i \partial_j \left( \frac{1}{R} \right) = -\frac{4\pi}{3} \, \delta_{ij} \delta(\boldsymbol{R}) + \frac{1}{R^3} \left( 3\tau_i \tau_j - \delta_{ij} \right) .$$

In the sense of generalized functions both kernels $R_{imln}$ and $S_{pqln}$ can be decomposed into a Dirac $\delta$-part and a $1/R^3$-part. Substituting Eq. (23) into Eqs. (13) and (17), we find

$$R_{imln} = \frac{1}{R^3} C_{jkln} H_{ijkm} + \delta(\boldsymbol{R}) \big[ \delta_{il} \delta_{mn} - C_{jkln} E_{ijkm} \big] \tag{29}$$

and

$$S_{pqln} = \frac{1}{R^3} C_{pqim} C_{jkln} H_{ijkm} + \delta(\boldsymbol{R}) \big[ C_{pqln} - C_{pqim} C_{jkln} E_{ijkm} \big] , \tag{30}$$

respectively. Using Eqs. (24) and (25), the kernel $R_{imln}$ reads explicitly for an arbitrary anisotropic medium [Lazar, 2016]

$$R_{imln} = \frac{1}{8\pi^2 R^3} \int_0^{2\pi} C_{jkln} \big[ 2\tau_m \tau_k \, (nCn)_{ij}^{-1} - 2(n_m \tau_k + n_k \tau_m) F_{ij} + n_m n_k A_{ij} \big] \mathrm{d}\phi$$
$$+ \delta(\boldsymbol{R}) \bigg( \delta_{il} \delta_{mn} - \frac{1}{2\pi} \int_0^{2\pi} C_{jkln} n_m n_k \, (nCn)_{ij}^{-1} \, \mathrm{d}\phi \bigg) , \tag{31}$$

which possesses a $1/R^3$-singularity and a Dirac $\delta$-singularity. Using Eqs. (24) and (25), the kernel $S_{pqln}$ reads explicitly for an arbitrary anisotropic medium [Lazar, 2016]

$$S_{pqln} = \frac{1}{8\pi^2 R^3} \int_0^{2\pi} C_{pqim} C_{jkln} \big[ 2\tau_m \tau_k \, (nCn)_{ij}^{-1} - 2(n_m \tau_k + n_k \tau_m) F_{ij}$$
$$+ n_m n_k A_{ij} \big] \mathrm{d}\phi + \delta(\boldsymbol{R}) \bigg( C_{pqln} - \frac{1}{2\pi} \int_0^{2\pi} C_{pqim} C_{jkln} n_m n_k \, (nCn)_{ij}^{-1} \, \mathrm{d}\phi \bigg) , \tag{32}$$

which possesses a $1/R^3$-singularity and a Dirac $\delta$-singularity.

## 3. Point defects

In the theory of anisotropic elasticity, point defects can be modeled as defects corresponding to a three-dimensional Dirac $\delta$-singularity in the eigendistortion tensor. The eigendistortion or quasi-plastic distortion tensor of an anisotropic point defect is given by

$$\beta_{ij}^{\mathrm{P}} = Q_{ij} \, \delta(\boldsymbol{R}) , \tag{33}$$

where $Q_{ij}$ is the strength of the point defect and $\boldsymbol{R} = \boldsymbol{x} - \boldsymbol{x}'$. In such a manner, a point defect is modeled as point defect with a Dirac $\delta(\boldsymbol{R})$-core and the specific character of the point defects is described by the explicit form of the tensor $Q_{ij}$ (see Table 1). The point defect corresponding to the eigendistortion (33) is located at point $\boldsymbol{x}'$. $Q_{ij}$ is called displacement dipole tensor [Kröner, 1956, 1958] and is, in general, an asymmetric polar tensor of rank two. The tensor $Q_{ij}$ has 9 independent components for arbitrary anisotropy (triclinic). The anisotropic tensor $Q_{ij}$ describes the character and type of the point defect under consideration. The symmetry and form of the second rank tensor $Q_{ij}$ is given in Table 1 for different crystal systems.

Table 1.   Symmetry and form of an asymmetric polar tensor $Q_{ij}$ of rank two (after Šhubnikov [1949] (see also Nye [1957]; Paufler [1986])).

| Symmetry | Crystal system | Form of $Q_{ij}$ | Orientation of axes |
|---|---|---|---|
| $\bar{1}$ | triclinic (both classes) | $\begin{pmatrix} Q_{11} & Q_{12} & Q_{13} \\ Q_{21} & Q_{22} & Q_{23} \\ Q_{31} & Q_{32} & Q_{33} \end{pmatrix}$ | arbitrary |
| $2/m$ | monoclinic (all classes) | $\begin{pmatrix} Q_{11} & Q_{12} & 0 \\ Q_{21} & Q_{22} & 0 \\ 0 & 0 & Q_{33} \end{pmatrix}$ | $2 \parallel x_3$ |
| $m\,m\,m$ | orthorhombic (all classes) | $\begin{pmatrix} Q_{11} & 0 & 0 \\ 0 & Q_{22} & 0 \\ 0 & 0 & Q_{33} \end{pmatrix}$ | $2 \parallel x_i$ |
| $\infty/m$ | trigonal: 3, $\bar{3}$ tetragonal: 4, $\bar{4}$, 4/m hexagonal: 6, $\bar{6}$, 6/m | $\begin{pmatrix} Q_{11} & Q_{12} & 0 \\ -Q_{12} & Q_{11} & 0 \\ 0 & 0 & Q_{33} \end{pmatrix}$ | $\infty \parallel x_3$ |
| $\infty/m\,m$ | trigonal: 32, 3m, $\bar{3}m$ tetragonal: 422, 4mm, $\bar{4}2m$, 4/mmm hexagonal: 622, 6mm, $\bar{6}m2$, 6/mmm | $\begin{pmatrix} Q_{11} & 0 & 0 \\ 0 & Q_{11} & 0 \\ 0 & 0 & Q_{33} \end{pmatrix}$ | $\infty \parallel x_3$ |
| $\infty\,\infty\,m$ | cubic (all classes) and isotropic | $\begin{pmatrix} Q_{11} & 0 & 0 \\ 0 & Q_{11} & 0 \\ 0 & 0 & Q_{11} \end{pmatrix}$ | arbitrary |

According to the theorem of Hermann [1934], a tensor of the rank $r$ has an $\infty$-fold symmetry axis if this tensor possesses an $N$-fold symmetry axis $N > r$. Thus, for a tensor of rank 2, the 3-fold, 4-fold and 6-fold symmetry axes are equal to an $\infty$-fold symmetry axis (see also Table 1).

Integrating the eigendistortion (33) over the volume $V$ containing the point defect, we get (see also deWit [1973b])

$$\int_V \beta_{ij}^{\mathrm{P}}\, \mathrm{d}V' = Q_{ij} \int_V \delta(\boldsymbol{R})\, \mathrm{d}V' = Q_{ij} \tag{34}$$

if $\boldsymbol{x}$ is in $V$. Here $\mathrm{d}V' = \mathrm{d}\boldsymbol{x}'$. Alternatively, the elastic dipole tensor or double force tensor [Kröner, 1958, 1981; Balluffi, 2012], which is a symmetric tensor of second rank, can be used and it is given by the symmetric part of the displacement dipole tensor for point defects according to

$$P_{ij} = C_{ijkl} Q_{kl}\,. \tag{35}$$

The quasi-dislocation density tensor (6) corresponding to the eigendistortion or quasi-plastic distortion (33) reads

$$\alpha_{ij} = -Q_{il}\epsilon_{jkl}\partial_k \delta(\boldsymbol{R})\,, \tag{36}$$

whereas $\partial_k \delta(\boldsymbol{R})$ represents the gradient of the three-dimensional Dirac $\delta$-function. Thus, the eigendistortion tensor (33) and the quasi-dislocation density tensor (36) possess a $\delta(\boldsymbol{R})$-singularity and a $\partial_k \delta(\boldsymbol{R})$-singularity, respectively. As known from field theory, the tensor (36) describes a "mathematical dipole" due to the term $\partial_k \delta(\boldsymbol{R})$.

Substituting the eigendistortion (33) into Eqs. (10), (14) and (16) and performing the convolution, we obtain the displacement field, total distortion tensor, elastic distortion tensor and stress tensor of an anisotropic point defect

$$u_i = -Q_{ln} C_{jkln} G_{ij,k} = -P_{jk} G_{ij,k} \tag{37}$$

$$\beta_{im}^{\mathrm{T}} = -Q_{ln} C_{jkln} G_{ij,km} = -P_{jk} G_{ij,km} \tag{38}$$

$$\beta_{im} = -R_{imln} Q_{ln} \tag{39}$$

$$\sigma_{pq} = -S_{pqln} Q_{ln} . \tag{40}$$

Now, using the explicit expression for the gradient of the Green tensor (20), the displacement field (37) of an anisotropic point defect in an anisotropic medium reduces to

$$\begin{aligned} u_i &= \frac{Q_{ln}}{8\pi^2 R^2} \int_0^{2\pi} C_{jkln} \big[ \tau_k (nCn)_{ij}^{-1} - n_k F_{ij} \big] \mathrm{d}\phi \\ &= \frac{P_{jk}}{8\pi^2 R^2} \int_0^{2\pi} \big[ \tau_k (nCn)_{ij}^{-1} - n_k F_{ij} \big] \mathrm{d}\phi . \end{aligned} \tag{41}$$

Using the kernels (31) and (32), Eqs. (38)–(40) and Eqs. (23)–(25), we obtain the following expressions for the elastic distortion of an anisotropic point defect in an anisotropic medium

$$\begin{aligned} \beta_{im} &= -\frac{Q_{ln}}{8\pi^2 R^3} \int_0^{2\pi} C_{jkln} \big[ 2\tau_m \tau_k \, (nCn)_{ij}^{-1} - 2(n_m \tau_k + n_k \tau_m) F_{ij} + n_m n_k A_{ij} \big] \mathrm{d}\phi \\ &\quad - \delta(\boldsymbol{R}) \, Q_{ln} \left( \delta_{il} \delta_{mn} - \frac{1}{2\pi} \int_0^{2\pi} C_{jkln} n_m n_k \, (nCn)_{ij}^{-1} \, \mathrm{d}\phi \right) \\ &= -\frac{P_{jk}}{8\pi^2 R^3} \int_0^{2\pi} \big[ 2\tau_m \tau_k \, (nCn)_{ij}^{-1} - 2(n_m \tau_k + n_k \tau_m) F_{ij} + n_m n_k A_{ij} \big] \mathrm{d}\phi \\ &\quad - \delta(\boldsymbol{R}) \left( Q_{im} - \frac{P_{jk}}{2\pi} \int_0^{2\pi} n_m n_k \, (nCn)_{ij}^{-1} \, \mathrm{d}\phi \right) , \end{aligned} \tag{42}$$

for the total distortion of an anisotropic point defect in an anisotropic medium

$$
\begin{aligned}
\beta_{im}^{\mathrm{T}} = &-\frac{Q_{ln}}{8\pi^2 R^3}\int_0^{2\pi} C_{jkln}\big[2\tau_m\tau_k\,(nCn)_{ij}^{-1} - 2(n_m\tau_k + n_k\tau_m)F_{ij} + n_m n_k A_{ij}\big]\mathrm{d}\phi \\
&+ \delta(\boldsymbol{R})\,\frac{Q_{ln}}{2\pi}\int_0^{2\pi} C_{jkln}n_m n_k\,(nCn)_{ij}^{-1}\,\mathrm{d}\phi \\
= &-\frac{P_{jk}}{8\pi^2 R^3}\int_0^{2\pi}\big[2\tau_m\tau_k\,(nCn)_{ij}^{-1} - 2(n_m\tau_k + n_k\tau_m)F_{ij} + n_m n_k A_{ij}\big]\mathrm{d}\phi \\
&+ \delta(\boldsymbol{R})\,\frac{P_{jk}}{2\pi}\int_0^{2\pi} n_m n_k\,(nCn)_{ij}^{-1}\,\mathrm{d}\phi
\end{aligned}
\tag{43}
$$

and for the stress of an anisotropic point defect in an anisotropic medium

$$
\begin{aligned}
\sigma_{pq} = &-\frac{Q_{ln}}{8\pi^2 R^3}\int_0^{2\pi} C_{pqim}C_{jkln}\big[2\tau_m\tau_k\,(nCn)_{ij}^{-1} - 2(n_m\tau_k + n_k\tau_m)F_{ij} \\
&+ n_m n_k A_{ij}\big]\mathrm{d}\phi - \delta(\boldsymbol{R})\,Q_{ln}\bigg(C_{pqln} - \frac{1}{2\pi}\int_0^{2\pi} C_{pqim}C_{jkln}n_m n_k\,(nCn)_{ij}^{-1}\,\mathrm{d}\phi\bigg) \\
= &-\frac{P_{jk}}{8\pi^2 R^3}\int_0^{2\pi} C_{pqim}\big[2\tau_m\tau_k\,(nCn)_{ij}^{-1} - 2(n_m\tau_k + n_k\tau_m)F_{ij} + n_m n_k A_{ij}\big]\mathrm{d}\phi \\
&- \delta(\boldsymbol{R})\bigg(P_{pq} - \frac{P_{jk}}{2\pi}\int_0^{2\pi} C_{pqim}n_m n_k\,(nCn)_{ij}^{-1}\,\mathrm{d}\phi\bigg).
\end{aligned}
\tag{44}
$$

It can be seen that the displacement vector (41) possesses a $1/R^2$-term similar to the electrostatic potential of an electric dipole and to the magnetostatic potential of a magnetic dipole (see Jackson [1999]; Griffiths [1999]), whereas the elastic distortion tensor (42), the total distortion tensor (43) and the stress tensor (44) possess $1/R^3$- and Dirac $\delta(\boldsymbol{R})$-terms. The $\delta(\boldsymbol{R})$-term does not contribute to the fields away from the position of the point defect and is called contact term. The $\delta(\boldsymbol{R})$-term is necessary to assure the solenoid character of the fields (42)–(44). Its purpose is to yield the required volume integral of those fields. Already Eshelby [1955, 1956] pointed out the importance of the $\delta(\boldsymbol{R})$-term in the dilatation field of a center of dilatation in a cubic material. Note that the distortion and stress fields (42)–(44) of a point defect are similar to the electric field of an electric dipole as well as to the magnetic field of a magnetic dipole consisting of a $1/R^3$-term and a $\delta(\boldsymbol{R})$-contact term (see Jackson [1999]; Griffiths [1999]; Frahm [1983]; Leung and Ni [2006]). From the field theoretical point of view, a point defect corresponding to the plastic distortion (33) represents an elastic dipole.

### 3.1. *Material force*

Consider the material force acting between defects. We derive the material force of a point defect in the stress field of another defect directly from the Peach-Koehler force. The Peach-Koehler force is defined by (see, e.g., Lazar and Kirchner [2013];

Agiasofitou and Lazar [2010]; Lazar [2016])

$$\mathcal{F}_s = \int_V \epsilon_{sjl} \sigma_{ij} \alpha_{il} \, \mathrm{d}V \,. \tag{45}$$

The Peach-Koehler force (45) is the interaction force between a dislocation density tensor $\alpha_{il}$ and a stress tensor $\sigma_{ij}$. Thus, the Peach-Koehler force is a configurational or material force acting between defects. Substituting the quasi-dislocation density tensor (36) of an anisotropic point defect into Eq. (45), using integration by parts and Eq. (1), we obtain the force exerted on the point defect in the gradient of a stress field

$$\mathcal{F}_s = Q_{pq} \sigma_{pq,s} \,. \tag{46}$$

Therefore, Eq. (46) has the physical interpretation as the interaction force between a point defect of strength $Q_{pq}$ and the gradient of a stress field $\sigma_{pq,s}$ which can be caused by other defects (point defect, dislocation). We note that Eq. (46) agrees with Kröner's force acting on a force dipole (see Kröner [1958, 1981]). On the other hand, the material force (46) can be expressed in terms of the elastic dipole tensor (35) and the gradient of the elastic strain tensor, $e_{ij} = 1/2(\beta_{ij} + \beta_{ji})$, (see also Kröner [1958, 1960])

$$\mathcal{F}_s = P_{ij} e_{ij,s} \,. \tag{47}$$

Using the stress tensor of a point defect (40), the interaction force (46) of a point defect with strength $Q_{pq}$ in the stress field of another point defect with strength $Q'_{ln}$ reads in terms of the gradient of the kernel $S_{pqln}$

$$\mathcal{F}_s = -Q_{pq} S_{pqln,s} Q'_{ln} \,, \tag{48}$$

where the gradient of the kernel (30) possesses $1/R^4$- and $\partial_s \delta(\boldsymbol{R})$-terms. The material force (48) is the force exerted by one point defect with strength $Q'_{ln}$ at $\boldsymbol{x}'$ on the other point defect with strength $Q_{pq}$ at $\boldsymbol{x}$. Here, $R$ is the distance between the two defects from $Q'_{ln}$ to $Q_{pq}$, and $\boldsymbol{R} = \boldsymbol{x} - \boldsymbol{x}'$ is the vector from $Q'_{ln}$ to $Q_{pq}$. Substituting the tensor (17) into Eq. (48), we obtain the interaction force between the two point defects

$$\mathcal{F}_s = -Q_{pq} C_{pqim} \big[ Q'_{ln} C_{jkln} G_{ij,kms}(\boldsymbol{R}) + Q'_{im} \partial_s \delta(\boldsymbol{R}) \big] \,. \tag{49}$$

Using the elastic dipole tensor (35), Eq. (49) simplifies to

$$\mathcal{F}_s = -P_{im} \big[ P'_{jk} G_{ij,kms}(\boldsymbol{R}) + Q'_{im} \partial_s \delta(\boldsymbol{R}) \big] \,. \tag{50}$$

Thus, the interaction force between two point defects consists of two contributions, namely a $1/R^4$-term and a $\partial_s \delta(\boldsymbol{R})$-term (contact term). The latter represents a short range force (contact force) between two point defects. Neglecting $\delta(\boldsymbol{R})$-terms, an expression of the third gradient of the Green tensor valid for $\boldsymbol{R} \neq 0$ has been given in [Goy *et al.*, 2009].

Consider the interaction force between a point defect at $\boldsymbol{x}$ in the stress field caused by a dislocation at $\boldsymbol{x}'$. Substituting the generalized Peach-Koehler stress formula of an arbitrary dislocation density $\alpha_{lr}$ [Lazar, 2016]

$$\sigma_{pq} = C_{pqim}\epsilon_{mnr}C_{jkln}G_{ij,k} * \alpha_{lr} \tag{51}$$

into Eq. (46), we obtain the interaction force between a point defect in the stress field of a dislocation

$$\mathcal{F}_s = Q_{pq}C_{pqim}\epsilon_{mnr}C_{jkln}G_{ij,ks} * \alpha_{lr} \, . \tag{52}$$

For a dislocation loop $C$, the dislocation density tensor reads [deWit, 1973a]

$$\alpha_{lr} = \oint_C b_l\delta(\boldsymbol{R})\, \mathrm{d}C_r' \, . \tag{53}$$

Inserting Eq. (53) into Eq. (52) yields the interaction force between a point defect and a dislocation loop

$$\mathcal{F}_s = \oint_C Q_{pq}b_lC_{pqim}\epsilon_{mnr}C_{jkln}G_{ij,ks}(\boldsymbol{R})\, \mathrm{d}C_r \, . \tag{54}$$

Using Eq. (23), the interaction force (54) reads

$$\mathcal{F}_s = \oint_C Q_{pq}b_lC_{pqim}\epsilon_{mnr}C_{jkln}\left(\frac{1}{R^3}\, H_{ijks} - \delta(\boldsymbol{R})\, E_{ijks}\right)\mathrm{d}C_r \, , \tag{55}$$

where the tensors $H_{ijks}$ and $E_{ijks}$ are given in Eqs. (25) and (24), respectively. The second term in Eq. (55) is a contact term.

## 3.2. *Interaction energy*

Consider the interaction energy between defects. The elastic interaction energy between an eigendistortion field $\beta_{pq}^{\mathrm{P}}$ and a stress field $\sigma_{pq}$ is given by [Mura, 1987]

$$U_{\mathrm{int}} = -\int_V \sigma_{pq}\beta_{pq}^{\mathrm{P}}\, \mathrm{d}V \, . \tag{56}$$

If we substitute the eigendistortion of a point defect (33) into Eq. (56) and carry out the integration, the interaction energy becomes

$$U_{\mathrm{int}} = -Q_{pq}\sigma_{pq} \, . \tag{57}$$

Eq. (57) represents the interaction energy of a point defect with strength $Q_{pq}$ in the stress field $\sigma_{pq}$ (see also Kröner [1958, 1981]). Comparing Eq. (46) with Eq. (57), we obtain the relation

$$\mathcal{F}_s = -\partial_s U_{\mathrm{int}} \, . \tag{58}$$

On the other hand, the interaction energy (57) can be expressed in terms of the elastic dipole tensor (35) and the elastic strain tensor (see also Kröner [1956, 1958, 1960]; Leibfried and Breuer [1978])

$$U_{\mathrm{int}} = -P_{ij}e_{ij} \, . \tag{59}$$

Using the stress tensor (40) of a point defect, the interaction energy of a point defect with strength $Q_{pq}$ in the stress field of another point defect with strength $Q'_{ln}$ can be written in terms of the kernel $S_{pqln}$.

$$U_{\mathrm{int}} = Q_{pq} S_{pqln} Q'_{ln}\,. \tag{60}$$

If we substitute Eq. (32) into Eq. (60), the explicit expression for the interaction energy between two point defects becomes

$$U_{\mathrm{int}} = \frac{Q_{pq} Q'_{ln}}{8\pi^2 R^3} \int_0^{2\pi} C_{pqim} C_{jkln} \big[ 2\tau_m \tau_k (nCn)_{ij}^{-1} - 2(n_m \tau_k + n_k \tau_m) F_{ij} + n_m n_k A_{ij} \big] \mathrm{d}\phi$$
$$+ \delta(\boldsymbol{R})\, Q_{pq} Q'_{ln} \bigg( C_{pqln} - \frac{1}{2\pi} \int_0^{2\pi} C_{pqim} C_{jkln} n_m n_k \, (nCn)_{ij}^{-1} \, \mathrm{d}\phi \bigg)\,. \tag{61}$$

The $\delta(\boldsymbol{R})$-term in Eq. (61) is a contact term in the interaction energy between point defects similar to the Fermi contact interaction of magnetic dipoles which is responsible for the hyperfine splitting of atomic spectra (see Jackson [1999]). If we use the elastic dipole tensor (35) and the tensors $H_{ijkm}$ and $E_{ijkm}$ given in Eqs. (25) and (24), respectively, then Eq. (61) can be written

$$U_{\mathrm{int}} = \frac{P_{im} P'_{jk}}{R^3}\, H_{ijkm} + \delta(\boldsymbol{R}) \big[ P_{ln} Q'_{ln} - P_{im} P'_{jk} E_{ijkm} \big]\,. \tag{62}$$

Eq. (62) represents the elastic interaction energy between two point defects for arbitrary anisotropy. The interaction energy (62) consists of two contributions, namely a $1/R^3$-term and a $\delta(\boldsymbol{R})$-term (contact term). In isotropic and anisotropic elasticity, contact terms are often erroneously ignored (see, e.g., Hardy and Bullough [1967]; Siems [1968]; Yoo [1974]; Schaefer and Kronmüller [1975]; Teodosiu [1982]).

Consider the interaction energy between a point defect in the stress field caused by a dislocation. Substituting the generalized Peach-Koehler stress formula (51) of an arbitrary dislocation density $\alpha_{lr}$ into the interaction energy expression (57) gives

$$U_{\mathrm{int}} = -Q_{pq} C_{pqim} \epsilon_{mnr} C_{jkln} G_{ij,k} * \alpha_{lr}\,. \tag{63}$$

Inserting Eq. (53) into Eq. (63) yields the interaction energy between a point defect with strength $Q_{pq}$ and a dislocation loop with Burgers vector $b_l$

$$U_{\mathrm{int}} = -\oint_C Q_{pq} b_l C_{pqim} \epsilon_{mnr} C_{jkln} G_{ij,k}\, \mathrm{d}C_r\,. \tag{64}$$

Moreover, using Eq. (20), the interaction energy (64) reads

$$U_{\mathrm{int}} = \oint_C \frac{Q_{pq} b_l}{8\pi^2 R^2}\, C_{pqim} \epsilon_{mnr} C_{jkln} \int_0^{2\pi} \big[ \tau_k (nCn)_{ij}^{-1} - n_k F_{ij} \big] \mathrm{d}\phi\, \mathrm{d}C_r\,. \tag{65}$$

Eq. (65) represents the elastic interaction energy between a point defect and a dislocation loop for arbitrary anisotropy.

### 3.3. *Elastic torque*

Consider the torque between defects. The elastic torque (or elastic rotational moment) between an eigendistortion field $\beta_{rq}^{\mathrm{P}}$ and a stress field $\sigma_{sq}$ is given by [Agiasofitou and Lazar, 2017]

$$\mathcal{T}_t = \int_V \epsilon_{trs}\beta_{rq}^{\mathrm{P}}\sigma_{sq}\,\mathrm{d}V\,. \tag{66}$$

If we substitute the eigendistortion of a point defect (33) into Eq. (66) and carry out the integration, the elastic torque becomes [Kröner, 1956]

$$\mathcal{T}_t = \epsilon_{trs}Q_{rq}\sigma_{sq}\,, \tag{67}$$

which is the interaction torque $\mathcal{T}_t$ exerted on the point defect with strength $Q_{rq}$ in a stress field $\sigma_{sq}$. It can be seen in Eq. (67) that an isotropic or cubic point defect, $Q_{rq} = \frac{1}{3}\delta_{rq}Q_{ll}$, does not produce a torque since the stress tensor is a symmetric tensor.

Using the stress tensor of a point defect (40), the torque of a point defect with strength $Q_{rq}$ due to the stress field of another point defect with strength $Q'_{ln}$ reduces to

$$\mathcal{T}_t = \epsilon_{tsr}Q_{rq}S_{sqln}Q'_{ln}\,. \tag{68}$$

If we substitute Eq. (32) into Eq. (68), the explicit expression for the elastic interaction torque between two point defects becomes

$$\mathcal{T}_t = \frac{Q_{rq}Q'_{ln}}{8\pi^2 R^3}\int_0^{2\pi}\epsilon_{tsr}C_{sqim}C_{jkln}\big[2\tau_m\tau_k(nCn)_{ij}^{-1} - 2(n_m\tau_k + n_k\tau_m)F_{ij} + n_mn_kA_{ij}\big]\mathrm{d}\phi$$
$$+ \delta(\boldsymbol{R})\,Q_{rq}Q'_{ln}\epsilon_{tsr}\left(C_{sqln} - \frac{1}{2\pi}\int_0^{2\pi}C_{sqim}C_{jkln}n_mn_k\,(nCn)_{ij}^{-1}\,\mathrm{d}\phi\right). \tag{69}$$

The $\delta(\boldsymbol{R})$ in Eq. (69) is a contact term. On the other hand, if we use the elastic dipole tensor (35) and the tensors $H_{ijkm}$ and $E_{ijkm}$ given in Eqs. (25) and (24), respectively, Eq. (69) can be written as

$$\mathcal{T}_t = \frac{1}{R^3}\epsilon_{tsr}Q_{rq}C_{sqim}P'_{jk}H_{ijkm} + \delta(\boldsymbol{R})\,\epsilon_{tsr}Q_{rq}\big[P'_{sq} - P'_{jk}C_{sqim}E_{ijkm}\big]\,. \tag{70}$$

Thus, the torque (70) between two point defects consists of two contributions, namely a $1/R^3$-term and a $\delta(\boldsymbol{R})$-term (contact term).

Consider the torque caused by a point defect in the stress field caused by a dislocation. Substituting the generalized Peach-Koehler stress formula (51) of an arbitrary dislocation density $\alpha_{lr}$ into the torque expression (67) yields

$$\mathcal{T}_t = \epsilon_{tps}Q_{pq}C_{sqim}\epsilon_{mnr}C_{jkln}G_{ij,k}*\alpha_{lr}\,. \tag{71}$$

Using Eq. (53), Eq. (71) gives the torque between a point defect with strength $Q_{pq}$ and a dislocation loop with Burgers vector $b_l$

$$\mathcal{T}_t = \oint_C \epsilon_{tps}Q_{pq}b_lC_{sqim}\epsilon_{mnr}C_{jkln}G_{ij,k}\,\mathrm{d}C_r\,. \tag{72}$$

Using Eq. (20), the torque (72) between a point defect and a dislocation loop is obtained

$$\mathcal{T}_t = -\oint_C \frac{Q_{pq}b_l}{8\pi^2 R^2} \epsilon_{tps} C_{sqim} \epsilon_{mnr} C_{jkln} \int_0^{2\pi} \left[ \tau_k (nCn)^{-1}_{ij} - n_k F_{ij} \right] \mathrm{d}\phi \, \mathrm{d}C_r \,. \qquad (73)$$

Eq. (73) represents the elastic torque between a point defect and a dislocation loop for arbitrary anisotropy.

Note that for a point defect in a homogeneous (constant) stress field, the interaction force is zero, $\mathcal{F}_s = 0$, whereas the interaction energy and the torque are non-zero, $U_{\mathrm{int}} \neq 0$ and $\mathcal{T}_t \neq 0$.

## 4. Eshelby factor meets Eshelby tensor

The so-called Eshelby factor plays an important role in the physics of point defects in crystals [Eshelby, 1954, 1956, 1975; Leibfried and Breuer, 1978; Maysenhölder, 1984; Michelitsch and Wunderlin, 1996]. The aim of this section is to connect the Eshelby factor with the Eshelby tensor.

Using the theory of (incompatible) anisotropic elasticity, the core of a point defect is modeled as a three-dimensional Dirac $\delta$-singularity (see Eq. (33)). Therefore, such a model is not appropriate for describing the detailed aspects of the structure of a point defect core, similar to the modeling of the dislocation core of a Volterra dislocation as two-dimensional Dirac $\delta$-singularity. However, within this model point defects can be connected to spherical inclusions (see also Balluffi [2012]).

The volume change due to the plastic (or quasi-plastic) dilatation of a point defect is given by

$$\Delta V = Q_{ii} = \int_V \beta_{ii}^{\mathrm{P}} \, \mathrm{d}V' \,, \qquad (74)$$

where $Q_{ii}$ represents the plastic volume change. The volume change due to the total dilatation of a point defect reads

$$\Delta V^\infty = \int_V u_{i,i} \, \mathrm{d}V' = \int_V \beta_{ii}^{\mathrm{T}} \, \mathrm{d}V' \,. \qquad (75)$$

$\Delta V^\infty$ gives the total change of a spherical reference volume of an anisotropic crystal, which is embedded in an infinite medium, under the influence of an anisotropic point defect. The volume change due to the negative elastic dilatation of a point defect is given by

$$\Delta V^I = -\int_V \beta_{ii} \, \mathrm{d}V' \,. \qquad (76)$$

In addition, it holds

$$\Delta V = \Delta V^\infty + \Delta V^I \,. \qquad (77)$$

In order to give the relation to micromechanics, we take the trace of the elastic distortion tensor in Eq. (12)

$$\beta_{ii} = -C_{jkln}G_{ij,ki} * \beta_{ln}^{\mathrm{P}} - \beta_{ii}^{\mathrm{P}}. \tag{78}$$

Using the decomposition (5), the trace of the total distortion tensor becomes

$$\beta_{ii}^{\mathrm{T}} = -C_{jkln}G_{ij,ki} * \beta_{ln}^{\mathrm{P}}. \tag{79}$$

We substitute the eigendistortion (33) into Eq. (79), perform the convolution, and we get

$$\beta_{ii}^{\mathrm{T}} = -Q_{ln}C_{jkln}G_{ij,ki}. \tag{80}$$

If we now substitute Eq. (80) into Eq. (75), we obtain

$$\Delta V^{\infty} = -Q_{ln}C_{jkln}\int_{V}G_{ij,ki}\,\mathrm{d}V'. \tag{81}$$

In order to perform the integration in Eq. (81), we may use the definition of the interior Eshelby tensor for a spherical inclusion in an anisotropic material (see, e.g., Li and Wang [2008]; Lazar [2016])

$$S_{imln}^{\mathrm{Esh}} = -\frac{1}{2}C_{jkln}\int_{V}\left[G_{ij,km}(\boldsymbol{R}) + G_{mj,ki}(\boldsymbol{R})\right]\mathrm{d}V'. \tag{82}$$

Note that the Eshelby tensor $S_{imln}^{\mathrm{Esh}}$ possesses the symmetries

$$S_{imln}^{\mathrm{Esh}} = S_{miln}^{\mathrm{Esh}} = S_{imnl}^{\mathrm{Esh}}. \tag{83}$$

The trace of the first two indices of the Eshelby tensor (82) gives

$$S_{iiln}^{\mathrm{Esh}} = -C_{jkln}\int_{V}G_{ij,ki}(\boldsymbol{R})\mathrm{d}V'. \tag{84}$$

Comparing Eq. (81) with Eq. (84), we observe that $\Delta V^{\infty}$ can be expressed in terms of the Eshelby tensor (84). Substituting Eq. (84) into Eq. (81), we find that the total volume change due to an anisotropic point defect in an anisotropic medium can be given in terms of the Eshelby tensor and the tensor $Q_{ln}$

$$\Delta V^{\infty} = S_{iiln}^{\mathrm{Esh}}Q_{ln}. \tag{85}$$

Now, the Eshelby factor for an anisotropic point defect in an anisotropic medium can be written in terms of the Eshelby tensor, namely

$$\gamma^{\mathrm{Esh}} = \frac{\Delta V}{\Delta V^{\infty}} = Q_{jj}\left[S_{iiln}^{\mathrm{Esh}}Q_{ln}\right]^{-1}. \tag{86}$$

Therefore, for an anisotropic point defect in an anisotropic medium the Eshelby factor depends on the Eshelby tensor and on the displacement dipole tensor $Q_{ij}$.

For a spherical inclusion in an anisotropic material, the interior Eshelby tensor is given by [Lazar, 2016]

$$S_{imln}^{\mathrm{Esh}} = \frac{1}{4\pi}C_{jkln}\int_{0}^{2\pi}\left[n_{m}n_{k}\,(nCn)_{ij}^{-1} + n_{i}n_{k}\,(nCn)_{mj}^{-1}\right]\mathrm{d}\phi. \tag{87}$$

Note that the anisotropic Eshelby tensor of a sphere (87) is given as line integral around the unit circle because it is based on the Lifshitz-Rosenzweig-Synge-Barnett representation of the anisotropic Green tensor (18). Taking the trace of the indices $i$ and $m$ of the Eshelby tensor (87)

$$S_{iiln}^{\text{Esh}} = \frac{1}{2\pi} C_{jkln} \int_0^{2\pi} n_i n_k \, (nCn)_{ij}^{-1} \, \mathrm{d}\phi \tag{88}$$

and substituting Eq. (88) into Eq. (86), the explicit form of the Eshelby factor of an anisotropic point defect in an anisotropic medium is obtained.

For a dilatational point defect (cubic or isotropic defect symmetry) the tensor $Q_{ij}$ reduces to

$$Q_{ij} = \frac{Q_{kk}}{3} \, \delta_{ij} \,, \tag{89}$$

where $Q_{kk} = 3Q_{11}$ (see Table 1). Combining Eqs. (86) and (89), the Eshelby factor for an isotropic or cubic point defect in an anisotropic medium reads

$$\gamma^{\text{Esh}} = 3 \big[ S_{iill}^{\text{Esh}} \big]^{-1} \,, \tag{90}$$

depending only on the trace of the indices $l$ and $n$ of the trace of the Eshelby tensor (88)

$$S_{iill}^{\text{Esh}} = \frac{1}{2\pi} C_{jkll} \int_0^{2\pi} n_i n_k \, (nCn)_{ij}^{-1} \, \mathrm{d}\phi \,, \tag{91}$$

which is the "double trace" of the Eshelby tensor (87). Due to the cubic defect symmetry, Eq. (90) delivers an analytical expression for the Eshelby factor for cubic crystals. Note that Maysenhölder [1984] gave the Eshelby factor for cubic crystals only in the Fourier space corresponding to the Fourier transform of Eq. (90).

Consider the special case of an isotropic material. The interior Eshelby tensor for a spherical inclusion in an isotropic material reads [Buryachenko, 2007; Li and Wang, 2008; Lazar, 2016]

$$S_{imln}^{\text{Esh}} = \frac{1}{15(1-\nu)} \left[ (5\nu - 1) \, \delta_{im} \delta_{ln} + (4 - 5\nu)(\delta_{il} \delta_{mn} + \delta_{ml} \delta_{in}) \right] \,, \tag{92}$$

where $\nu$ is the Poisson ratio. Taking the trace in the indices $i$ and $m$

$$S_{iiln}^{\text{Esh}} = \frac{1+\nu}{3(1-\nu)} \, \delta_{ln} \,. \tag{93}$$

If we substitute Eq. (93) into Eq. (85), then the volume change due to the total dilatation of a point defect in an isotopic material reads

$$\Delta V^\infty = \frac{1+\nu}{3(1-\nu)} \, Q_{ll} \,. \tag{94}$$

Finally, using the isotropic Eshelby tensor (93) and Eqs. (86) and (94), we recover the original Eshelby factor for an isotropic medium given by Eshelby [1954, 1956]

$$\gamma^{\text{Esh}} = \frac{\Delta V}{\Delta V^\infty} = \frac{3(1-\nu)}{1+\nu} \,, \tag{95}$$

depending only on the Poisson ratio. Note that the Eshelby factor (95) is valid for isotropic point defects in an isotropic medium as well as for anisotropic point defects in an isotropic medium since only the trace part $Q_{ll}$ gives a contribution.

## 5.  Conclusions

In this work, we derived key-equations of point defects in anisotropic elasticity from the perspective of micromechanics. Point defects have been modeled as $\delta$-type singularities in the eigendistortion. We derived the material force (interaction force), interaction energy and elastic torque, in general, and applied to the interaction of two point defects as well as of a point defect and a dislocation loop. In particular, the interaction force, interaction energy and torque between two point defects are given in terms of the kernel $S_{pqln}$. Similarities between point defects and electric and magnetic dipoles are pointed out; especially so-called contact terms. In addition, the Eshelby factor for anisotropic point defects in arbitrary anisotropic crystals has been derived in terms of the Eshelby tensor. It is worth noticing that the derived key-equations of point defects are easy to implement into numerical codes, since the appearing integrals are "well-behaved" functions. The derived point defect formulation is a contribution to micromechanics in general, which may have impact to discrete dislocation dynamics, defect mechanics, lattice theory of defects and computational engineering of defects.

## Acknowledgements

The author gratefully acknowledges grants from the Deutsche Forschungsgemeinschaft (Grant Nos. La1974/3-2, La1974/4-1). The author is grateful to Thomas Michelitsch (Paris) for useful remarks. The author also wishes to thank two anonymous reviewers for their encouragement and comments.